\documentclass[twocolumn,amssymb,amsmath,aps,secnumarabic,floatfix]{revtex4}
\usepackage{psfig}
\usepackage{epsfig}
\expandafter\ifx\csname package@font\endcsname\relax\else
 \expandafter\expandafter
 \expandafter\usepackage
 \expandafter\expandafter
 \expandafter{\csname package@font\endcsname}%
\fi
\begin{document}
\title{Broken scaling in the Forest Fire Model}
\author{Gunnar Pruessner and Henrik Jeldtoft Jensen}
\affiliation{
Department of Mathematics,
Imperial College,
180 Queen's Gate,
London SW7 2BZ,
UK\\
gunnar.pruessner@physics.org
and
h.jensen@ic.ac.uk
}
\date{\today}

\begin{abstract}
We investigate the scaling behavior of the cluster size distribution in
the Drossel-Schwabl Forest Fire model (DS-FFM) by means of large
scale numerical simulations, partly on (massively) parallel machines. 
It turns out that simple scaling is clearly violated, as
already
pointed out by Grassberger 
{\bracketOpen}P. {Grassberger}, {J. Phys. A: Math. Gen.} \textbf{26},
 {2081} ({1993}){\bracketClose}, 
but largely
ignored in the literature. Most surprisingly the statistics not
 seems to be described by a universal scaling function, and the scale of
 the physically relevant region seems to be a constant.
 Our results strongly suggest that the DS-FFM is not
critical in the sense of being free of characteristic scales.
\end{abstract}
\maketitle

\section{Introduction}
The Drossel-Schwabl Forest Fire Model \cite{DrosselSchwabl:1992}
(DS-FFM) is one of the paradigms of
non-conservative SOC \cite{Jensen:98}. Its
importance comes primarily from the fact that the model has
non-conservative microdynamics. It therefore answered the question
whether conservation is necessary for criticality in driven systems.

The claim that the DS-FFM is critical comes from the fact that it shows
powerlaw-like behavior for several geometrical properties of the
dissipation events. The average size of these is divergent in the
so-called SOC limit, where all timescales get separated so that the rate
of the external drive becomes infinitely slow and the total inflow
diverges. If one assumes stationarity this is trivial to prove 
\cite{DrosselSchwabl:1992,ClarDrosselSchwabl:1994,ClarDrosselSchwabl:1996}.
However, as usual in numerical simulations, it is not  
possible to investigate the model in the limit of divergent drive
($\theta \to 0$ in the notation below), as
finite size limits 
the correlation length and therefore destroys any possible criticality
\cite{ClarDrosselSchwabl:1994}.  
It is remarkable that most of the literature available for this
model is mainly concerned with finding critical exponents and
identifying supposedly critical quantities. It seems that no authors
question whether the model is critical at all and if so in which
sense. In this paper we carefully investigate the ``scaling function'' of
the cluster size distribution and show that it is indeed an open
question whether the model is truly critical: Not only is there no
way to prove its criticality, there is also numerical evidence that the
model does not become scale free.

\section{Definition of the model and methods}
\label{sec:methods}
The model has been described several times and in great detail elsewhere
\cite{DrosselSchwabl:1992,ClarDrosselSchwabl:1996,ClarDrosselSchwabl:1994}.
Therefore, the description presented here is rather succinct.
The model is defined on a d-dimensional lattice of linear size
$L$, where each lattice site can be in one of two states,
``occupied'' or ``empty''. The lattice is then updated according to the
following rules: 
\begin{itemize}
\item Driving:
      Choose randomly $\INFL$ sites, one after the other. If its state
      is ``empty'' turn it into ``occupied''.
\item Relaxation:
      Choose one site at random. If it is empty continue with the first
      step. Otherwise make ``empty'' all the sites in the cluster the
      site chosen belongs to. In this case
      the update is considered to be successful. Continue with the
      first step.
\end{itemize}
Here a cluster is defined in the usual fashion as the set of occupied
sites which are connected via nearest neighbor interactions, i.e. two
sites belong to the same cluster if they are nearest neighbors or there
is a path between them along sites, which belong to the same cluster. We
have applied periodic 
boundaries in all our simulations and restrict ourselves to the
two dimensional square lattice.

The cluster removed in the second step is called the ``burnt
cluster''. To measure the overall distribution of clusters within the
system, one usually measures the size of the burnt cluster
\cite{ClarDrosselSchwabl:1994}, the distribution of which is biased by a
factor $s$. To see that, we define 
$n(s)$ to be the ensemble-averaged, site-normalized density of
clusters of size $s$ in 
the system. Then, the probability that a randomly chosen site is
connected to a cluster of size $s$ is $s n(s)$, as in standard
percolation \cite{StaufferAharonyENG:1994}. This distribution is probably
the most important in the model. Other quantities are the distribution
of the burning time, which is defined as the maximum Manhattan distance
(shortest path on the square lattice) from the initially chosen site of a 
burnt cluster to all other sites in the same cluster, and the correlation functions as
defined and discussed in \cite{HoneckerPeschel:1997}. In this paper we
solely concentrate on the distribution $n(s)$.

Using a new implementation of the model
\cite{JensenPruessnerFFM_technique_TBP} we are able to simulate the 
system on very large scales and at the same time keep track of the
{\it entire} distribution $n(s)$, instead of measuring the biased
distribution $s n(s)$, as done usually
\cite{ClarDrosselSchwabl:1994}. Between two 
updates the changes in $n(s)$ are only of the order $\INFL$, so it is
a highly correlated quantity. However, by using standard cluster labeling
techniques \cite{HoshenKopelman:1976}, it is possible to calculate
the full histogram $n(s)$ essentially without increasing the computing
time, which depends almost exclusively on $\theta$ and is essentially
independent of the system size. Compared to the standard simulation, we
gain up to two orders of 
magnitude in performance \footnote{
As the standard deviation (actually the estimator of the standard
deviation of the estimated mean; this quantity includes the
correlation time) decreases with the square root of the computing
time, one has to compare the products of the square root of the computing
time and the standard deviation. In the algorithm presented here, the
ratio of the computing time varies between $1.4$ and $2.1$
}. A similar method was recently introduced for standard percolation
\cite{NewmanZiff:2000}. 
Using the same amount of computing time the results are
significantly less noisy than those of the standard 
implementation (for example \cite{HoneckersFFMCode}, which we have used
as reference to check the validity of our results).
Large system sizes enable us to rule out any finite size effects as
described below. The results
have been partly cross-checked using a different random number generator
(all results presented here make use of {\tt ran2} from \cite{Press:92},
and for checking {\tt ran1} from \cite{Press:92} has been used).

Finite size effects have been ruled out by the following direct method:
For each 
value of $\INFL$ a significantly larger system was simulated with exactly
the same value for $\INFL$. The linear size $L$ was typically increased
by a factor $2$. The smallest systems we used were $L=1000$, the largest
$L=32000$. By comparing the histograms of different system sizes in
conjunction with the standard deviation calculated for them, it is
possible to decide whether a system size is affected by finite size effects
or not. Compared to other simulations
published \cite{Schenk:2001,Schenk:2000}, which also claim not to suffer
from finite size effects, our system sizes are huge. This simple method of
``redoing'' all simulations and using lattices which are much larger than
actually needed has the obvious disadvantage of being inefficient, but
there is probably no more a direct way of 
identifying finite size effects \cite{SchLoiPru:2001}. This waste of
computing power is overcompensated by the efficiency of the algorithm
and self-averaging \cite{FeLaBi:91}.

\section{Results}
The focus of this paper is the presumably universal scaling function of
the distribution $n(s)$. Similar to the correlation function one expects
\begin{equation}
 n(s; \theta) = s^{-\tau} \GC(s/s_0(\theta) )
\label{eq:simple_scaling}
\end{equation}
if ``simple scaling'' applies, which is already known not to be the case
in the presence of finite size effects \cite{Schenk:2000}. The
$L$-dependence of this quantity is omitted in the following wherever the
context allows it. It is
worthwhile to note that this 
is usually the {\it definition} of the exponent $\tau$. The function
$\GC$ is the (presumably) universal scaling function, which depends
only on the ratio $s/s_0(\theta)$, where $s_0(\theta)$ is the {\it only} scale of the
distribution. This scale depends only on the external parameters, in our
case $\theta$. Simple scaling allows another scale, namely the lower
cutoff, but this is fixed or at least bounded. 

The function $\GC(x)$ is usually smooth for small values of $x$,
therefore it does not make a big difference whether we investigate
$n(s)$ or 
\begin{eqnarray}
 \RN(s;\theta) & = & \frac{n(s;\theta)}{n(1;\theta)} \nonumber \\
  & = & \frac{\GC(s/s_0(\theta))}{\GC(1/s_0(\theta))} s^{-\tau}  \quad .
\end{eqnarray}
That this particular choice of the normalization does not affect the
overall results can be seen in Tab.~\ref{tab:absolute_results}, where
the absolute value of $n(1,\theta)$ is listed for different values of
$\theta$. Also shown in this table is the first moment of the distribution
or the average density of trees, which is defined as 
\begin{equation}
 \rho = \sum_{s} s n(s;\theta)
\end{equation}
and the second moment of the distribution, 
\begin{equation}
 \ave{s} = \frac{\sum_{s} s^2 n(s;\theta)}{\sum_{s} s n(s;\theta)} \quad,
\label{eq:ave_s}
\end{equation}
which is the average size of the cluster connected to a randomly chosen
occupied site.

\begin{table}
\caption{\label{tab:absolute_results}
Static quantities for different choices of $L$ and $\INFL$. The
 estimation of the standard deviation of the tree density $\rho$,
 $\sigma^2(\rho)=\ave{\rho^2}-\ave{\rho}^2$, where the average runs over
 the ensemble, is
 unfortunately based only on a small subset of the configurations
 produced, and in the case of the large systems ($L\ge 16000$) only on a
 fraction of the lattice. However, it is apparent that it behaves like
 $1/L$, as expected for a system without finite size effects. The
 density of clusters of size $1$, $n(1)$, serves as the normalization of
 $\RN$. The average cluster size is denoted by $\ave{s}$, for definition
 see (\ref{eq:ave_s}), but due to a truncation in the histogram for some
 of the simulations in the range $2000\le \INFL \le 16000$, the
 number presented is actually the average size of the burnt cluster. In
 the stationary state it is - apart from statistical fluctuations - also given by 
 $(1-\ave{\rho})/(\theta \ave{\rho})$ \cite{ClarDrosselSchwabl:1994}.
Values of $\INFL$ and $L$
 printed in bold indicate results shown in Fig.~\ref{fig:scale_collapse},
 the other results are only for comparison. 
All data are based on $5\cdot 10^6$ (successful) updates
 (s. Sec.\ref{sec:methods}) for the transient and statistics, apart from
 those printed in italics which are based on short runs ($5\cdot 10^6$
 updates for the transient and $1\cdot 10^6$
 updates for statistics).
}
\begin{tabular}{r|r|r|r|r|r|r}
$\INFL$    & $L$        & $\ave{\rho}$    & $\sqrt{\sigma^2(\rho)}$ &
 $n(1)$      & $\ave{s}$    & $\frac{1-\ave{\rho}}{\theta \ave{\rho}}$ \\
\hline 
{\it 125}    & {\it 1000}  & 0.3797       & 0.0060                  & 0.04553     & 204.07     & 204.18  \\
     125     &      1000   & 0.3798       & 0.0058                  & 0.04552     & 203.81     & 204.15   \\
{\it 125}    & {\it 4000}  & 0.3798       & 0.0014                  & 0.04553     & 203.88     & 204.10  \\
{\bf 125}    & {\bf 4000}  & 0.3798       & 0.0015                  & 0.04552     & 203.77     & 204.10   \\
{\it 250}    & {\it 1000}  & 0.3876       & 0.0083                  & 0.04451     & 395.03     & 395.06  \\
     250     &      1000   & 0.3875       & 0.0082                  & 0.04452     & 394.08     & 395.15   \\
{\it 250}    & {\it 4000}  & 0.3877       & 0.0022                  & 0.04454     & 394.97     & 394.89  \\
{\bf 250}    & {\bf 4000}  & 0.3877       & 0.0021                  & 0.04454     & 395.29     & 394.91   \\
{\it 500}    & {\it 1000}  & 0.3932       & 0.0117                  & 0.04380     & 764.73     & 771.75  \\
     500     &      1000   & 0.3932       & 0.0119                  & 0.04380     & 764.81     & 771.77 \\
{\it 500}    & {\it 4000}  & 0.3934       & 0.0031                  & 0.04382     & 771.12     & 770.88  \\
{\bf 500}    & {\bf 4000}  & 0.3934       & 0.0030                  & 0.04382     & 771.90     & 770.87  \\
{\it 1000}   & {\it 1000}  & 0.3972       & 0.0169                  & 0.04328     & 1495.36   & 1517.91  \\
     1000    &      1000   & 0.3971       & 0.0168                  & 0.04328     & 1490.05   & 1518.00    \\
{\it 1000}   & {\it 4000}  & 0.3976       & 0.0043                  & 0.04331     & 1510.85   & 1515.00   \\
{\bf 1000}   & {\bf 4000}  & 0.3976       & 0.0043                  & 0.04331     & 1513.13   & 1514.81    \\
{\it 1000}   & {\it 8000}  & 0.3976       & 0.0021                  & 0.04332     & 1510.10   & 1514.91   \\
{\it 2000}   & {\it 4000}  & 0.4005       & 0.0060                  & 0.04296     & 2976.34    & 2993.35  \\
{\bf 2000}   & {\bf 4000}  & 0.4005       & 0.0062                  & 0.04297     & 2990.50    & 2993.15   \\
{\it 2000}   & {\it 8000}  & 0.4006       & 0.0030                  & 0.04297     & 2995.67    & 2992.56  \\
{\it 4000}   & {\it 4000}  & 0.4026       & 0.0089                  & 0.04273     & 5929.24     & 5935.91 \\
     4000    &      4000   & 0.4025       & 0.0089                  & 0.04273     & 5930.97     & 5938.03  \\
{\it 4000}   & {\it 8000}  & 0.4026       & 0.0048                  & 0.04274     & 5931.32     & 5935.15 \\
{\bf 4000}   & {\bf 8000}  & 0.4026       & 0.0046                  & 0.04273     & 5935.36     & 5936.47  \\
{\it 8000}   & {\it 4000}  & 0.4040       & 0.0135                  & 0.04255     & 11786.97    & 11799.72 \\
     8000    &      4000   & 0.4041       & 0.0135                  & 0.04255     & 11788.90    & 11799.07  \\
{\it 8000}   & {\it 8000}  & 0.4041       & 0.0068                  & 0.04257     & 11801.31    & 11795.98 \\
{\bf 8000}   & {\bf 8000}  & 0.4041       & 0.0068                  & 0.04257     & 11792.82    & 11795.38 \\
{\it 16000}  & {\it 4000}  & 0.4052       & 0.0199                  & 0.04244     & 23430.01    & 23481.82 \\
{\it 16000}  & {\it 8000}  & 0.4054       & 0.0096                  & 0.04243     & 23466.93    & 23467.22 \\
     16000   &       8000  & 0.4054       & 0.0098                  & 0.04243      &  23446.10   & 23465.64   \\
{\bf 16000}  & {\bf 16000} & 0.4054       & 0.0052                  &  0.04245      & 23449.31   & 23466.57   \\
     32000   &      16000  & 0.4066       & 0.0075                  & 0.04232     &  46443.83    & 46701.82 \\
{\bf 32000}  & {\bf 32000} & 0.4066       & 0.0032                  &  0.04233    &   46731.44   & 46698.51 \\
{\bf 64000}  & {\bf 32000} & 0.4078       & 0.0042                  &  0.04220     &   91148.64  & 92952.40 \\
\end{tabular}
\end{table}

Before presenting the actual results, we first discuss the
numerical quality of the results. 

\subsection{Avoiding finite size effects}
\begin{figure}[th]
\begin{center}
      \includegraphics[width=0.9\linewidth]{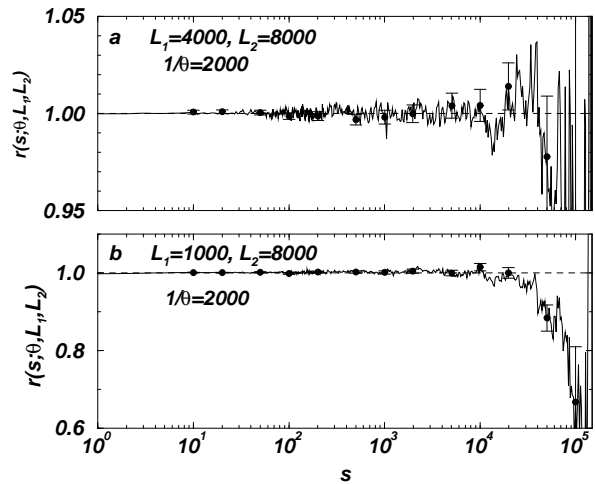}
\caption{\label{fig:ffs_check} Ratio $r(s;\theta, L_1, L_2) = \RN(s;\theta, L_1)/\RN(s;\theta, L_2)$ 
with $\INFL=2000$ for two pairs $L_1, L_2$ with error-bars (one standard
 deviation; the error-bars as well as the data shown are exponentially
 binned). The data are from short runs ($10^6$ updates for
 statistics). Finite size effects have been considered negligible under
 the condition that (almost all) 
 error-bars for this ratio have covered $1$ (marked by a dashed line) in
 the relevant range. 
 a) $L_1=4000$ and $L_2=8000$: Almost no finite size effects, the
 deviation from $1$ is probably due to noise. Note the
 fine scale of the ordinate. b) $L_1=1000$ and $L_2=8000$: Systematic,
 strong finite size effects for $s \gtrapprox 10^4$. The scale of the ordinate is five times
 larger than in a). Data of this quality have been dismissed.}
\end{center}
\end{figure}
\begin{figure}[th]
\begin{center}
      \includegraphics[width=0.9\linewidth]{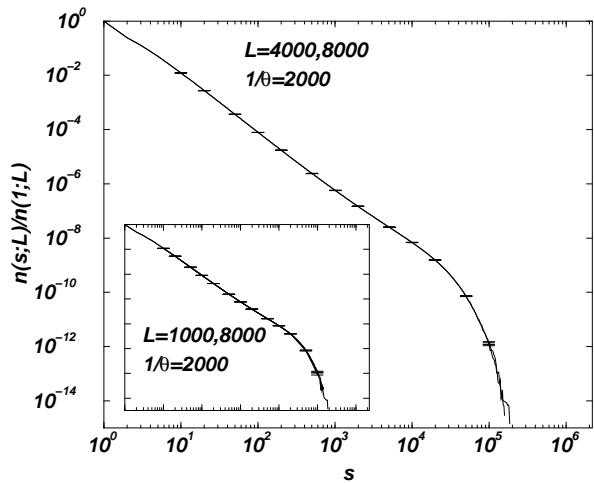}
\caption{\label{fig:ffs_check_visual} The binned histogram
 $\RN(s;\theta,L)$ for two different values of $L$ and fixed $\theta$ as
 in Fig.~\ref{fig:ffs_check}a.  
 In this plot the two histograms are virtually indistinguishable.
 However, note that the deviations shown on Fig.~\ref{fig:ffs_check}b would also
 hardly be visible in this type of plot, as shown in the inset.
}
\end{center}
\end{figure}

Throughout this paper we initially performed $5\cdot 10^6$
successful updates (as defined in Sec.~\ref{sec:methods}) as transient
(and therefore rejected them) and the same 
number for producing statistics, apart from runs for calculating
error-bars, where only $10^6$ updates has been used for statistics, see
below. 
It is known that the transient can be very long
\cite{HoneckerPeschel:1997} (note that the time unit in \cite{HoneckerPeschel:1997} 
is expressed in our units by multiplying it with  $\INFL / (\rho L^2)$), but in all
cases presented the number of initial steps seemed to be more than
sufficient. Numerical checks indicate that the
cluster size distribution is very stable against the size of the
transient, i.e. even a transient, which is presumably too short, still produces
reasonable results for $n(s)$.

All systems have been initialized by a random independent distribution of
trees with density $0.41$.

The standard deviation of the binned histogram is not completely trivial
to calculate. In particular, its computation requires a significant amount
of CPU time, and was therefore only calculated for the smaller
system sizes (up to $L=8000$) and in shorter runs (only $10^6$ updates
for statistics, but $5\cdot 10^6$ for transient). We resorted to visual
examination for the larger systems when comparing $\RN(s;\theta,L)$ for
different system sizes. 
Fig.~\ref{fig:ffs_check}a and b show the ratio of $\RN(s;\theta,L)$ for two different
system sizes. A deviation of this ratio from $1$ indicates a difference
in the statistics and therefore the presence of finite size effects.
Fig.~\ref{fig:ffs_check}a shows a typical case we accepted as reasonable
agreement. Here $L_1=4000$ and $L_2=8000$ do not seem to differ
for $\INFL=2000$. Fig.~\ref{fig:ffs_check}b shows a case of finite size
corrections we have dismissed (note the different scales in the
two graphs). It differs from Fig.~\ref{fig:ffs_check}a only by $L_1=1000$.

Fig.~\ref{fig:ffs_check_visual} illustrates the strong
agreement of $\RN(s;\theta)$ at the same value of $\theta$ for the same two
different sizes $L$ as in Fig.~\ref{fig:ffs_check}a. The two sets of
data are virtually indistinguishable, but in this kind of plot it is also
almost impossible to see a difference between the data of $L_1=1000$ and
$L_2=8000$, as shown in the inset of
Fig.~\ref{fig:ffs_check_visual}. This is also the case with the
rescaled data below, and the use of very large systems throughout
this paper might therefore be 
``overcautious'' in avoiding finite size effects, although such
large sizes are obviously required for an accurate \emph{quantitative} analysis of
this model. However, when 
it comes only to qualitative analysis, such a judgment seems to be
justified. On the other hand, an increase in system size hardly
increases the computing time and affects ``only'' the memory requirements,
which forced us to implement the algorithm for parallel
machines. The side effect of using multiple CPUs at the same time is a
significant reduction of the simulation time especially for large values of
$\INFL$, a fact which compensates the complications of parallel
coding.

Another indicator for the absence of finite size effects is the scaling
of the standard deviation of $\rho$: If the lattice can be split into
independent parts, i.e. if subsets of the lattice can be considered as
independent, the standard deviation of $\rho$ should scale like
$1/L$ for different values of $L$ at given $\INFL$. Such a behavior can
be seen in Tab.~\ref{tab:absolute_results}, although the standard
deviation of $\rho$ could be calculated only roughly. This might explain
the slight mismatch for $\INFL=32000$, $L=16000,32000$.

For the highest values of $\INFL$ we could not yet do the comparison to
another system, so the curve for the largest value of $\INFL$ in
Fig.~\ref{fig:scale_collapse} is dotted, as their quality is not
known. However, it is reasonable to
assume that it is not affected by finite size scaling.

\subsection{The scaling function}
\begin{figure}[th]
\begin{center}
      \includegraphics[width=0.9\linewidth]{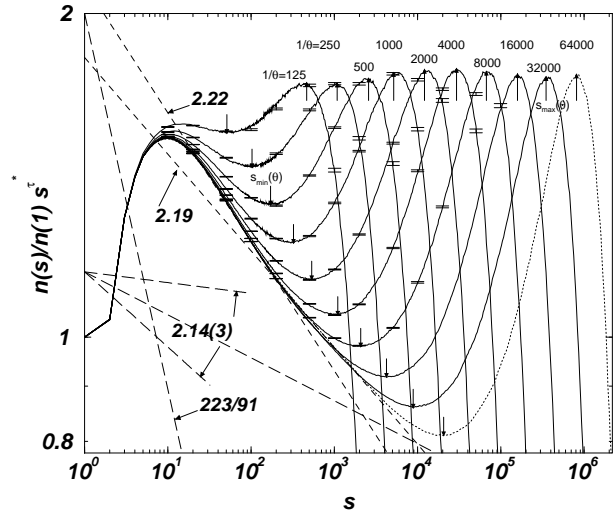}
\caption{\label{fig:scale_collapse} The rescaled and binned histogram
 $\RN(s;\theta) s^{\TR}$, where $\TR=2.10$ for 
$\INFL=125, 250, 500, \cdots, 32000, 64000$ (as indicated) in a double logarithmic
plot. The linear size $L$ is chosen according to the bold printed
 entries in Tab.~\ref{tab:absolute_results} and large enough to ensure
 absence of finite size 
 effects. The error-bars are estimated from shorter runs. The rightmost
histogram (in all figures dotted, $\INFL=64000$) could not be crosschecked by another
run, see text. Maxima are marked by arrows pointing upwards, minima are
 marked by arrows pointing downwards. The dashed lines belong to 
different exponents, whose value is specified as the sum of the slope in the
 diagram and $\TR$, i.e. a horizontal line would correspond to an exponent
 $2.1$. The shortly dashed line are estimated exponents for
different regions of the histogram ($2.22$ within approx. $[20,200]$
 and  $2.19$ within $[200,2000]$), the other exponents are from
literature, namely $2.14(3)$ in
 \cite{ClarDrosselSchwabl:1994,ClarDrosselSchwabl:1996} and
 $223/91\approx 2.45$ in \cite{Schenk:2001}. Since it was impossible
 to relate these exponents to  any property of the data, the exact
 position of the lines associated with them was chosen arbitrarily.
}
\end{center}
\end{figure}

\begin{figure}[th]
\begin{center}
      \includegraphics[width=0.9\linewidth]{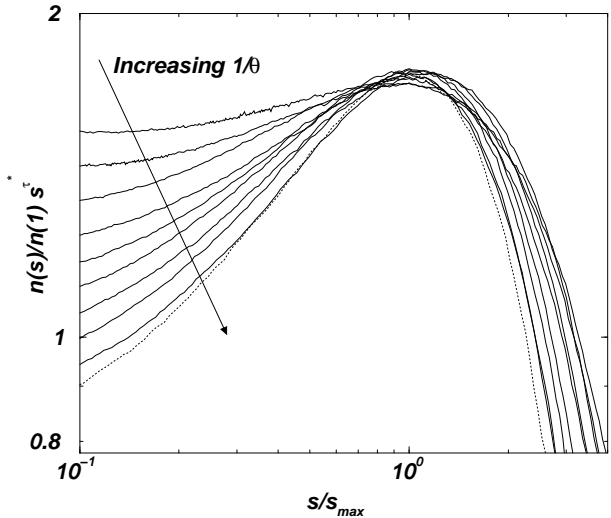}
\caption{\label{fig:second_bump_scaling} The rescaled and binned 
 histogram $\RN(s;\theta) s^{\TR}$, versus $s/s_{\ST{max}}(\theta)$, where $\TR=2.10$ for
 $\INFL=125, 250, 500, \cdots, 32000, 64000$ in a double logarithmic
 plot. The scales $s_{\ST{max}}(\theta)$ by which the histograms have been shifted
 are the maxima marked in Fig.~\ref{fig:scale_collapse}, so that a data
 collapse would be possible. 
The arrow indicates the order of the data in increasing $\INFL$.
}
\end{center}
\end{figure}

Comparing the different histograms $\RN(s;\theta)$ for different values
of $\INFL$ in a plot enables us not only to find the
exponent $\tau$, but also to find the universal function $\GC$ as
defined in Eqn.~\ref{eq:simple_scaling}. A rough, naive estimate of
$\tau$ is given by $\RN(s;\theta)$ fitted against $s^{-\tau}$, which
gives a value of $\TR \approx 2.1$ in our case. Plotting
now $\RN(s;\theta) s^{\TR}$ double
logarithmically should allow us to find the ``true'' value of $\tau$ by
performing a data collapse, i.e. choosing $\TR$ in such a way that
horizontal shifts (corresponding to the choice of the scale $s_0(\theta)$
in the scaling function) make all curves collapse. This is shown in
Fig.~\ref{fig:scale_collapse}, where $\TR=2.1$ was chosen so that the
maxima for the second bumps are almost equally high: denoting their
position on the abscissa for each value of $\theta$ by
$s_{\ST{max}}(\theta)$, we have chosen $\TR$ such that
\begin{equation}
 \RN(s_{\ST{max}}(\theta);\theta)\; s_{\ST{max}}^{\TR}(\theta)
\approx \text{const.} \quad .
\end{equation}
According to (\ref{eq:simple_scaling}) the constant is simply the
maximum value of $\GC$, namely $\GC(s_{\ST{max}}(\theta) / s_0(\theta)
)$, where the value of the argument is therefore the same for all
$\theta$. 

The value of $\TR$ is close to (but not within the error of)
the exponent found in the literature, $\tau=2.14(3)$
\cite{ClarDrosselSchwabl:1994,ClarDrosselSchwabl:1996} ($\tau=2.15(2)$ in
\cite{Grassberger:1993}, $\tau=2.159(6)$ in \cite{HoneckerPeschel:1997}),
which is shown in the same figure for comparison. 
However, it is impossible to force the minima (see the down pointing
marks in Fig.~\ref{fig:scale_collapse}) to the same height while
maintaining the constraint that the maxima remain aligned,
i.e. these minima cannot be a feature of the same universal
scaling function. Otherwise (\ref{eq:simple_scaling}) would hold and the
quantity  
\begin{equation}
  \RN(s_{\ST{min}}(\theta);\theta)\; s_{\ST{min}}^{\TR}(\theta)
\quad ,
\end{equation}
where $s_{\ST{min}}(\theta)$ denote the position of the minima,
would assume the same value for all $\theta$, because they are local
minima of $\GC$, which is supposed to be the same for all $\theta$.

Since these minima cannot be included in the simple scaling defined in
(\ref{eq:simple_scaling}), they 
must be explicitly excluded by introducing a lower cutoff, so that simple scaling
supposedly sets in only above these cutoffs, excluding especially the
minima. However, such a lower cutoff would apparently have to
diverge for $\INFL \to \infty$ -- something that is certainly
beyond any established concept of scaling. Even when accepting this
peculiar scaling behavior, a data collapse for the
second bump still seems to be unsatisfactory, as shown in
Fig.~\ref{fig:second_bump_scaling}. 

If one accepts a divergent lower cutoff of the scaling function, one has
to face the fact that this would describe the behavior of $\RN$ in a
region, which becomes physically less and less interesting in the
limit $\INFL \to \infty$, because the vast majority of events are situated at
small $s$ and as the second bump moves out to infinity, the scaling
function hence covers a smaller and smaller part of $\RN$. However, 
only a region of $\RN$ which covers a non-vanishing fraction of events
can be {\it physically relevant}.

Concentrating now on the behavior of $\RN$ up to the minimum (see
arrows pointing downwards in Fig.~\ref{fig:scale_collapse}), one finds
that this region is also badly described by a function like
(\ref{eq:simple_scaling}). First of all,
the question of which region is supposedly described by the function
needs to be answered. A unique lower cutoff and a $\theta$ dependent
upper cutoff needs to be found. At first view it looks appealing to
choose these two marks such that they cover the set of data, where the
curves fall on top of each other. In this case the lower cutoff would be
$1$ and the upper cutoff, $s_{\text{{\small naive}}}$, would have a value
smaller than the minima marked by downwards pointing
arrows in Fig.~\ref{fig:scale_collapse}. 
However, this would be described by a function like
\begin{equation}
  \RN(s;\theta) = f(s) \GC(s/s_{\text{{\small naive}}})
  \label{eq:early_collapse}
\end{equation}
rather than (\ref{eq:simple_scaling}). Note the {\it parameter
independent} function $f(s)$ describing the shape of the curve, while
$\GC(s/s_{\text{{\small naive}}})$ is a sharp cutoff
function. Eqn.~\ref{eq:early_collapse} does 
not allow for an exponent, $f(s)$ is an arbitrary function. Writing it
as
\begin{equation}
 f(s) = s^{-\tau} (a_0 + \text{higher order corrections})
\end{equation}
defines $\tau$ to be the steepest descent of this part of the curve and
gives a value between $\tau_{\ST{stp.}}=2.22$ and  $\tau_{\ST{stp.}}=2.19$
(see Fig.~\ref{fig:scale_collapse}). 

\begin{figure}[th]
\begin{center}
      \includegraphics[width=0.9\linewidth]{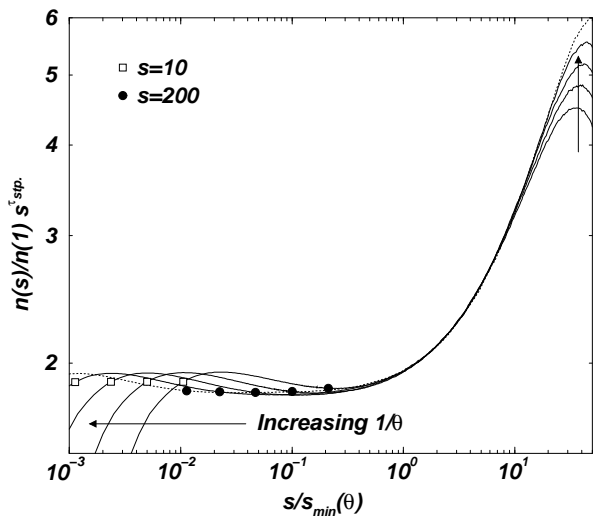}
\caption{\label{fig:early_scaling} The rescaled and binned 
 histogram $\RN(s;\theta) s^{\tau_{\ST{stp.}}}$, versus $s/s_{\ST{min}}(\theta)$, 
 where $\tau_{\ST{stp.}}=2.19$ for
 $\INFL=4000, 8000, 16000, 32000, 64000$ in a double logarithmic
 plot. The scales $s_{\ST{min}}(\theta)$ by which the histograms have been shifted
 are slightly different from the minima marked in
 Fig.~\ref{fig:scale_collapse}, to make the collapse as good as
 possible. The squares and the filled circles mark $s=10$ and
 $s=200$, respectively, for orientation and relation to other figures.
The arrows indicate the order of the data in increasing $\INFL$.
}
\end{center}
\end{figure}

This concept appears to be rather naive -- on the other hand, it is hard
to assume that (\ref{eq:simple_scaling}) can still hold: it would
correspond to (\ref{eq:early_collapse}) with $f(s)$ replaced by
$s^{\tau}$, which is a straight line in a double logarithmic
plot. Therefore (\ref{eq:simple_scaling}) can apply only to a region
in Fig.~\ref{fig:scale_collapse} where the data that fall on top of each
other form a straight line. Those features not already collapsing would
then collapse when properly tilted (choosing the right $\tau$) and shifted
(choosing the right $s_0$). Introducing a lower cutoff at $s=10$ and
discarding the data for $\INFL \le 2000$ then leads to a
data collapse in a narrow range as shown in
Fig.~\ref{fig:early_scaling}. It is worthwhile mentioning that even for
some $10<s<200$, namely for values of $s$ between the squares and
the filled circles, none of the data collapse. The exponent used in this
``collapse'' is $\tau_{\ST{stp.}}=2.19$, as mentioned above.

By considering the function $f(s)$ it becomes apparent that $\RN$, and
therefore the model, cannot be scale free: it depends on the
fixed, microscopic scale 
$s=1$. This entails that it is always possible to tell $\INFL$ by
looking only at the {\it shape} of $\RN$; a diagram showing only this
shape, without any scales on the axes, reveals $\INFL$, since a scale is
intrinsically given by the features of $f(s)$. One would only need to
rescale and tilt it until it fits the plot
Fig.~\ref{fig:scale_collapse} and one could identify $\INFL$. Only if
$f(s)$ were scale free, i.e. a straight line in a double logarithmic
plot, would this not be possible.

\begin{figure}[t]
\begin{center}
      \includegraphics[width=0.9\linewidth]{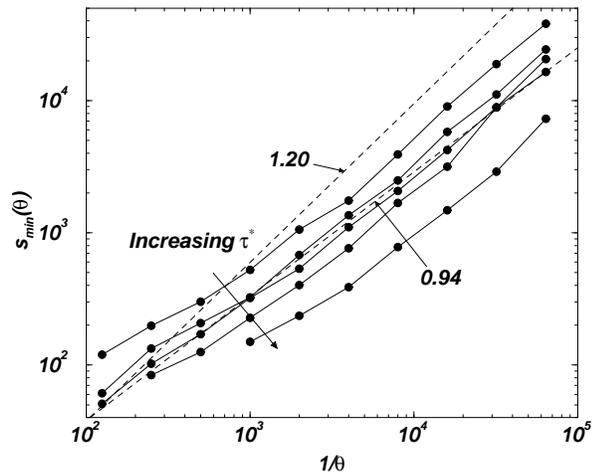}
\caption{\label{fig:min} The position of the minimum in the binned and
 rescaled histogram for different values of $\TR=2.04, 2.08, 2.10, 2.12, 2.16$. 
The exponents shown in the plot are for comparison only.
}
\end{center}
\end{figure}
\begin{figure}[t]
\begin{center}
      \includegraphics[width=0.9\linewidth]{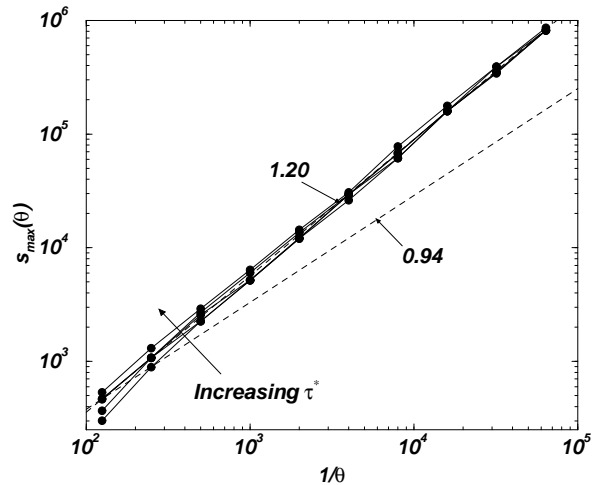}
\caption{\label{fig:max} The position of the maximum in the binned and
 rescaled histogram for different values of $\TR=2.04, 2.08, 2.10, 2.12, 2.16$. 
The exponents shown in the plot are for orientation only.
}
\end{center}
\end{figure}

\subsection{Two length scales}
That $\RN$ contains features to define at least two scales, which apparently
diverge in $\INFL$ with different exponents, becomes clear when analyzing the
scaling of the minima and maxima as marked in
Fig.~\ref{fig:scale_collapse}, using the definitions
\begin{eqnarray}
 s_{\ST{min}}(\theta) & \propto & \INFL^{x_{\ST{min}}} \\
 s_{\ST{max}}(\theta) & \propto & \INFL^{x_{\ST{max}}} \quad .
\end{eqnarray}
Of course, the exact position of the extrema of $\RN(s;\theta) s^{\TR}$ depends on
its tilt, i.e. on the choice of $\TR$. However, their \emph{scaling}
in $\INFL$ does not depend strongly on this choice. In particular
$x_{\ST{min}}$ and $x_{\ST{max}}$ are different for all choices of $\TR$. A plot of
$s_{\ST{min}}(\theta)$ versus $\INFL$ for different values
of $\TR$ is shown in Fig.~\ref{fig:min}. For small values of
$\INFL$ the minimum is not pronounced enough to survive for large values
of $\TR$, so these curves do not give a data point. Using a linear fit
of $\log{s_{\ST{min}}(\theta)}$ versus $\log{\INFL}$ of the minimum as found in the
rescaled ($\TR$) and binned histogram, gives an ``exponent'' between
$x_{\ST{min}}=0.93$ and $x_{\ST{min}}=0.98$. The same procedure applied
to the maxima gives an ``exponent'' in the range $x_{\ST{max}}=1.18$ and
$x_{\ST{max}}=1.22$, shown in Fig.~\ref{fig:max}. One may expect that
$x_{\ST{min}}$ tends towards $x_{\ST{max}}$ for decreasing $\TR$, as
$s_{\ST{min}}$ increases and might enter the scaling region of
$s_{\ST{max}}$, but neither ``exponent'' exhibits
a systematic variation, and the 
quality of the fit certainly suffers from the rough procedure that searches for
the extrema in the \emph{binned} histogram. This is unfortunately
necessary because of statistical fluctuations, in conjunction with the
absence of error-bars for all data points.

The scale of the clusters, $s_{\ST{min/max}}$ is related to the correlation
length $\xi$ by the fractal dimension $\mu$, i.e. (see \cite{ClarDrosselSchwabl:1994})
\begin{equation}
 s_{\ST{min/max}} \propto \xi^{\mu_{\ST{min/max}}} \quad .
\end{equation}
Since $\xi \propto \INFL^\nu$, one should expect $\nu =
x_{\ST{min/max}}/\mu_{\ST{min/max}}$. The minima   
are supposed to be dominated by smaller, fractal events (see \cite{Schenk:2001}), so
$\mu_{\ST{min}}=1.96(1)$ \cite{ClarDrosselSchwabl:1994} and 
therefore $\nu_{\ST{min}}\in [0.47,0.50]$. The maxima are more likely
to be dominated by compact fires, so $\nu_{\ST{max}} \in [0.59,0.61]$. 
It is unclear how the two exponents $\nu_{\ST{min/max}}$ are related
exactly to the exponents of the two
correlation lengths found by Honecker and Peschel
\cite{HoneckerPeschel:1997} for the connected 
correlation function $\nu=0.576(3)$ and for the tree-tree correlation
function $\nu=0.541(4)$.

\begin{figure}[t]
\begin{center}
      \includegraphics[width=0.9\linewidth]{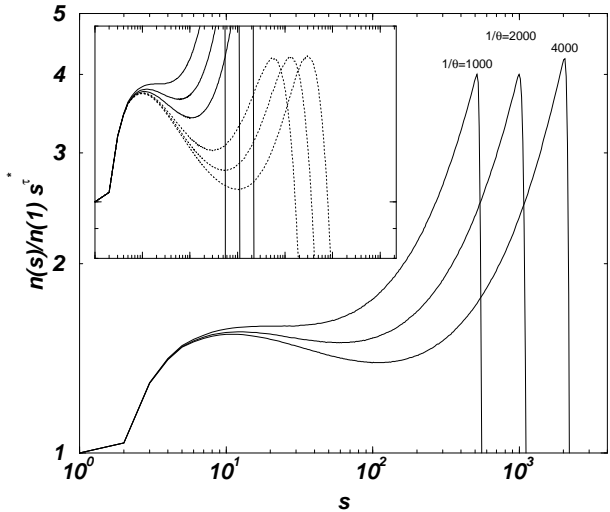}
\caption{\label{fig:ffm_extremal_drive}  The rescaled and binned
 histogram $\RN(s;\theta) s^{\TR}$ (again $\TR=2.10$), for a modified
 model, where the largest cluster in the system is removed after each
 driving step, for $\INFL=1000,2000,4000$ (as indicated) with linear sizes
 $L=2000,2000,4000$. The inset shows the same data on the scale of
 Fig.~\ref{fig:scale_collapse} for comparison. The data for
 $\INFL=1000,2000,4000$ of the original model as shown in
 Fig.~\ref{fig:scale_collapse} are dotted. The peculiar behavior
 of the different height-scaling of the minimum and the maximum is again visible.
}
\end{center}
\end{figure}
\begin{figure}[t]
\begin{center}
      \includegraphics[width=0.9\linewidth]{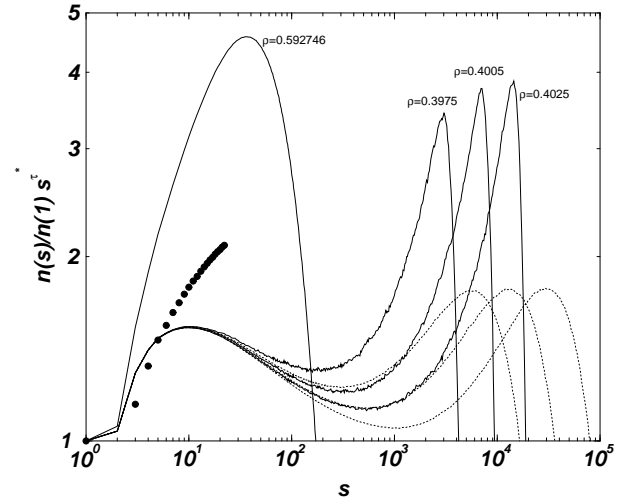}
\caption{\label{fig:ffm_perc_drive} 
The rescaled and binned
 histogram $\RN(s;\rho) s^{\TR}$ (again $\TR=2.10$), for a modified
 model, where the largest cluster in the system is removed in each
 relaxation step and the corresponding number of trees is filled back
 into the system afterwards.
 The three small values of $\rho$ chosen, $\rho=0.3975, 0.4005, 0.4025$
 correspond (up to the last digit) to the values of the tree density for
 $\INFL=1000,2000,4000$ respectively, see
 Tab.~\ref{tab:absolute_results}. The linear size was
 $L=1000,2000,4000$. The corresponding data of the original
 model are shown dotted.
 The peculiar behavior of the different height-scaling of the minimum
 and the maximum is again visible (a correct tilt $\TR$ would make it
 even more pronounced), but disappears obviously for
 $\rho=\rho_{\text{perc}}$ -- for these data it is relevant to mention
 that $\RN(s)$ was measured {\it after} the relaxation. The filled
 circles show the exact results for the lattice animals 
 \cite{StaufferAharonyENG:1994,SykesGlen:1976,Mertens:1990}
 at $\rho=\rho_{\text{perc}}$.
}
\end{center}
\end{figure}

\section{Discussion}
{\it Prima facie} the DS-FFM looks like a percolation process, and one
might naively think that it is indeed a percolation process which
organizes itself to the critical density: sites are occupied randomly
and independently and (at least in the thermodynamic limit) there is only
one cluster which is removed with non-vanishing probability, namely the
largest. In this way the density of occupied sites is automatically reduced
below the percolation threshold whenever the threshold is reached. It is
puzzling how remarkably close
the tree density in the DS-FFM is to the density of {\it empty} sites in
critical site percolation on a square lattice 
($\rho_{\text{FFM}}\approx 0.4078$ and $1-\rho_{\text{perc}}=0.40725379(13)$
\cite{NewmanZiff:2000} respectively). 
However, the removal process involved in the DS-FFM introduces
spatial correlations which are not present in standard percolation. These
correlations are expressed, for example, in the form of a patchy tree density
distribution \cite{Schenk:2001}.

The purpose of this paper is \emph{not} to add yet another model to the
enormous zoo of SOC models. However, in order to investigate certain
features of the given model and identify underlying mechanisms, it makes
sense to modify it slightly. 
The outcome for the histogram of the DS-FFM modified such that the
\emph{largest} cluster is removed after each driving step is shown for a few
values of $\INFL$ in Fig.~\ref{fig:ffm_extremal_drive}. The distinctive
feature of a minimum which scales differently from the maximum is again
present, as the peaks of the maxima have approximately the same
height, while the height of the local minima varies among different
values of $\theta$. The inset of this figure shows the histogram on the same scale
as Fig.~\ref{fig:scale_collapse} together with the data of the original
model (dotted) with the corresponding values of $\theta$. One can
understand that they do not fall on top of each other, because the
relaxation rule in the modified model erases much larger clusters than in
the original model. 

Fig.~\ref{fig:ffm_perc_drive} shows a second 
modification of the model, where again the {\it largest} cluster is
removed during relaxation and in addition the driving is changed such
that the density of trees, $\rho$, is the same before each
relaxation; the trees removed during the relaxation are just filled in
randomly afterwards. This model differs from standard percolation only
by its updating scheme \footnote{Actually, it also differs from standard
percolation because it fixes the number of occupied sites rather
than simply the probability of being occupied. However, this difference
becomes irrelevant for sufficiently large systems.}. In order to compare 
the outcome with the original model, the values of $\rho$ have been chosen close
to the values given in Tab.~\ref{tab:absolute_results}. Indeed, the feature
of different scaling of the extrema is still present, but it disappears
completely if the density is 
increased to $\rho_{\text{perc}}=0.592746$ \cite{NewmanZiff:2000}, which is shown in the same figure as
the large bump. This curve does not vary much if a much smaller
system size is simulated at this density, so we expect it essentially to be
free of finite size corrections. Since it represents a \emph{correlated}
percolation process, it is just consistent that
this bump does not cover the exact results for the lattice animals 
of standard percolation
\cite{StaufferAharonyENG:1994,SykesGlen:1976,Mertens:1990} 
at $\rho=\rho_{\text{perc}}$ shown as filled circles in 
Fig.~\ref{fig:ffm_perc_drive}.
The dotted graphs in the figure show the corresponding data of the
original model. Again they do not match apart from the region of very
small $s$. Unfortunately the simulations of the so-modified model are 
very expensive in CPU time, because the mass of the largest cluster
needs to be refilled after each relaxation, so that only
$50.000$ updates for transient and statistics could be done.

Since the feature of different scaling survives the modifications described
above, it seems reasonable to assume that any relaxation rule that
favors the largest cluster leads to the peculiar behavior. Its
disappearance at high densities can be explained by the extremely small
cutoff in the distribution, which leads to a domination of the
statistics by very small 
clusters, while a single, enormously large one dominates the burning
(the average size of the burnt cluster for $\rho=\rho_{\text{perc}}$ was
$355811$). However, much more careful and detailed investigations of
models like the modification described above are required to gain a full
understanding of the underlying mechanisms. In particular, this should 
include a modification of the rules such that the feature disappears.

Honecker and Peschel \cite{HoneckerPeschel:1997} have calculated the
correlation length not only for the probability that two sites belong
to the same cluster, but also for the probability that two sites are
occupied at all. The correlation function for the latter is of course a
$\delta$ peak in ordinary percolation, as there are no spatial correlations
for the distribution of occupied sites by construction. However, in the
DS-FFM the correlation length for this quantity, $\xi$, is finite and seems to diverges when
approaching the critical point. It is highly remarkable that Honecker
and Peschel
conclude from their simulations that this correlation length diverges
slightly {\it slower } than the correlation length of the probability for two
sites to belong to the same cluster, $\xi_s$. This seems to indicate that for
sufficiently large scales the spatial correlation of the occupation
probability becomes arbitrarily small, so that on sufficiently large scales
the DS-FFM occupation is uncorrelated and therefore tends to standard
percolation. In other words, it seems to be possible to rescale or ``renormalize'' 
the DS-FFM to make the occupation correlation arbitrarily small. This would
introduce higher order interactions, as known from standard real space
renormalization group and would explain the difference in critical
density between the rescaled DS-FFM and standard percolation. However,
if this ``mapping'' is valid, one should 
find the exponent for the divergence of $\xi_s/\xi$ to be the same as in
standard percolation, but this is precluded by numerics.

It has been suggested at least twice
\cite{HoneckerPeschel:1997,Schenk:2001}, that the DS-FFM is a superposition
of cluster distributions $n_{\ST{perc}(s,p)}$ of standard percolation for a
whole range of 
concentrations $p$, weighted by a certain distribution function $w(p)$, i.e.
$\int_0^1 dp\, w(p) n(s,p)$. Obviously such an assumption neglects
spatial correlations. 
We recall the following result from standard percolation theory
\cite{StaufferAharonyENG:1994},
\begin{equation}
 n(s,p) \propto s^{-\tau} \CC(-s/(p-p_c)^{-1/\sigma}) \quad ,
\label{eq:standard_perc_scaling}
\end{equation}
where $\CC$ denotes the cutoff function and the exponents $\sigma$ and
$\tau_{\ST{perc}}$ have their standard definitions. Assuming that the
weighting function $w(p)$ is analytic around  
the critical concentration in standard percolation, $p_c$,
(\ref{eq:standard_perc_scaling}) leads
to 
\begin{equation}
 \int_0^1 dp w(p) n(s,p) \propto s^{-(\tau_{\ST{perc}} + \sigma)} \quad .
\end{equation}
This gives rise to an exponent $\tau = 223/91 \approx 2.45$, however,
this is definitely not supported by numerics (see Fig.~\ref{fig:scale_collapse}).

It remains completely unclear how to characterize the scaling of the
DS-FFM in two dimensions. Apparently it is not a mere
superposition of two simple
scalings, as recently speculated \cite{Schenk:2001}. Moreover the model
does not seem to be scale free as described above and it does not seem
to be possible to identify a unique power law behavior of the cluster
size distribution. Nevertheless \emph{effective} power law behavior
over restricted regions has clearly been produced by the model, making
it potentially relevant to observation.

All we can conclude is
that \emph{the DS-FFM is not critical} in the sense of simple
scaling. It
reminds us that a divergent moment (here $\ave{s}$, the second moment) can be
regarded as a unique sign of emergent scale invariance only if
we are certain that one single scale is sufficient to characterize the
system. If there is more than one relevant scale, different
properties of the system might depend on  different scales which may or
may not diverge.

\begin{acknowledgments}
The large scale simulations represent the backbone of this paper. Andy
Thomas at the Department of Mathematics at Imperial College was eager to
support us technically, making machines available and upgrading the
systems in our department such that parallel computing became a
pleasure. A great deal of the results in this paper was possible only
because of his work. We thank him very much. 

This work partly relies on resources provided by the Imperial College
Parallel Computing Centre. We want to especially thank K. M. Sephton for
his support.

Another part of this work was possible only because of the generous
donation made by ``I-D Media AG, Application Servers \& Distributed
Applications Architectures, Berlin''. We especially thank M. Kaulke
and O. Kilian for their support. 

G.P. wishes to thank A. Honecker, I. Peschel and K. Schenk for
 very helpful communication, as well as N. R. Moloney for providing the
 lattice animal data and for proofreading.

The authors gratefully acknowledge the support of EPSRC.

\end{acknowledgments}
\bibliography{articles,books}
\end{document}